\documentclass[aps,prl,reprint]{revtex4-1}
\usepackage{blindtext}
\usepackage{amsmath}
\usepackage{amssymb}
\usepackage{braket}
\usepackage{graphicx}
\usepackage{physics}
\usepackage{color}
\usepackage{fancyhdr}
\usepackage{siunitx}
\usepackage{subcaption}
\usepackage[font=small,labelfont=bf]{caption}
\usepackage[colorinlistoftodos]{todonotes}

\begin{document}
\title{Phase stability transfer across the optical domain using a commercial optical frequency comb system}
\author{Lee Peleg}
\email{lee.peleg@weizmann.ac.il}
\author{Nitzan Akerman}
\author{Tom Manovitz}
\author{Meir Alon}
\author{Roee Ozeri}
\affiliation{Department of Physics of Complex Systems, Weizmann Institute of Science, Rehovot 7610001, Israel}

\begin{abstract}
We report the frequency noise suppression of a \SI{674}{\nano\meter} diode laser by phase-locking it to a \SI{1560}{\nano\meter} cavity-stabilized laser, using a commercial optical frequency comb. By phase-locking the frequency comb to the narrow reference at telecom wavelength we were able to phase-coherently distribute the reference stability across the optical spectrum. Subsequently, we used one of the comb teeth as an optical reference for a \SI{674}{\nano\meter} external cavity diode laser. We demonstrated the locked \SI{674}{\nano\meter} laser frequency stability by comparing it to an independent cavity-stabilized laser of the same wavelength and by performing spectroscopic measurements on a dipole-forbidden narrow optical transition in a single $^{88}$Sr$^+$ ion. These measurements indicated a fast laser-linewidth of \SI{19}{\hertz} and \SI{16}{\hertz}, respectively.
\end{abstract}
\maketitle
\section{introduction}
 Narrow linewidth laser systems, exhibiting high frequency stability, are an important tool in many scientific fields, such as precision spectroscopy and optical atomic clocks\cite{ludlow2015optical,nicholson2012comparison,hinkley2013atomic}, tests of fundamental physics\cite{chou2010optical,eisele2009laboratory} and gravity wave detection\cite{abbott2016observation}. \\
 Over the years, several technologies and methods were developed for the generation of stable optical references. The most common scheme involves phase-locking a laser source to a high-finesse and ultra stable optical cavity using electronic feedback in a Pound-Drever-Hall configuration\cite{Drever1983}. Such laser systems were shown to exhibit fractional frequency instabilities of a few times $10^{-16}$ in seconds of integration \cite{Swallows2012,Alnis2008,B.C.YoungW.M.Itano1999,Jiang2011}, ultimately limited by the thermal instability of the cavity length \cite{Kessler2011}. Advanced cavities, developed to deliver enhanced frequency stability, are often wavelength-specific and cumbersome to setup \cite{Kessler2012}. Being so, the motivation to distribute the phase coherence of a cavity-stabilized laser to other lasers in different domains of the optical spectrum arises naturally.\\
 The transfer of stability between two narrow-linewidth lasers was demonstrated with the utilization of an optical frequency comb as a coherent bridge over the optical spectral gap \cite{Hagemann2013,Yamaguchi2012,nicolodi2014spectral,Scharnhorst2015, Fang2015,akamatsu2013spectroscopy,Inaba2013,Akamatsu:12,yamanaka2015frequency,nemitz2016frequency,yamaguchi2011direct}. The heart of the scheme relies on exploiting the correlations between distinct discrete frequency modes of the frequency comb, usually referred to as frequency teeth. The relation between the comb repetition rate $f_r$, the carrier to envelope phase shift $\Delta\phi_{ceo}$ and the n\textsuperscript{th} comb mode $\nu_n$\cite{Ye2005} is,
 \begin{align}
 \begin{split}
 \nu_n =  n\cdot f_r + \frac{\Delta\phi_{ceo}}{2\pi}f_r \equiv n\cdot f_r + f_{ceo}.
 \end{split}
 \end{align}
 It therefore follows that by phase locking $\nu_n$ to a spectrally narrow optical reference, such as a cavity-stabilized clock laser, each other comb mode $\nu_m$ will inherit the master laser phase noise spectrum, with its power scaled by $\left(\frac{m}{n}\right)^2$; This is provided that the lock is sufficiently fast, the carrier-envelope offset frequency is independently stabilized and any superimposed extra-cavity noise has sufficiently low power\cite{newbury2007low}. Early frequency combs had limited servo bandwidth on the repetition frequency, resulting in poor tracking of the optical frequency reference at high frequencies. Therefore, in early demonstrations of this method, other techniques were incorporated to combat the fast phase noise of the locked comb modes, most commonly pre-stabilization of the target laser to an external optical cavity \cite{Hagemann2013,Yamaguchi2012}  or establishing a fast feed-forward scheme \cite{stenger2002ultraprecise} to eliminate the correlated comb noise \cite{Grosche2008, nicolodi2014spectral}. This scheme was also used with high bandwidth frequency combs, due to excessive phase-noise at high-Fourier components of each comb mode \cite{Scharnhorst2015}.   Using custom built combs, exhibiting high servo bandwidth, low residual phase noise and sub-Hz linewidths\cite{Fang2013Optical,nakajima2010multi,ohmae2017all}, high bandwidth phase stability transfer for locking ECDL lasers was demonstrated \cite{Fang2015,akamatsu2013spectroscopy, Inaba2013,Akamatsu:12, yamanaka2015frequency, nemitz2016frequency}.\\
 In this work we use a low-noise commercial comb system (Menlo Systems GmbH; FC1500-250-ULN) to transfer the phase stability of a commercially available telecom laser (Stable Laser Systems; SLS-INT-1550-200-1) to an external cavity diode laser operating at a wavelength of \SI{674}{\nano\meter}. By comparing this laser against a different, cavity-stabilized, laser at \SI{674}{\nano\meter} we were able to estimate the laser fast-linewidth to be $<$\SI{20}{Hz}, reaching a fractional frequency stability of $<3\cdot 10^{-15}$ (corresponding to sub-\si{\hertz} drifts) after \SI{1}{\second}. We later Used this laser for a spectroscopic measurement of the $5S_{1/2} \to 4D_{5/2}$ optical clock transition in \textsuperscript{88}Sr\textsuperscript{+} trapped ion, as a further examination of the stabilized laser spectral properties.
 
\section{experiment}
\begin{figure*}[t]
	\centering
	\includegraphics[width=0.8\textwidth]{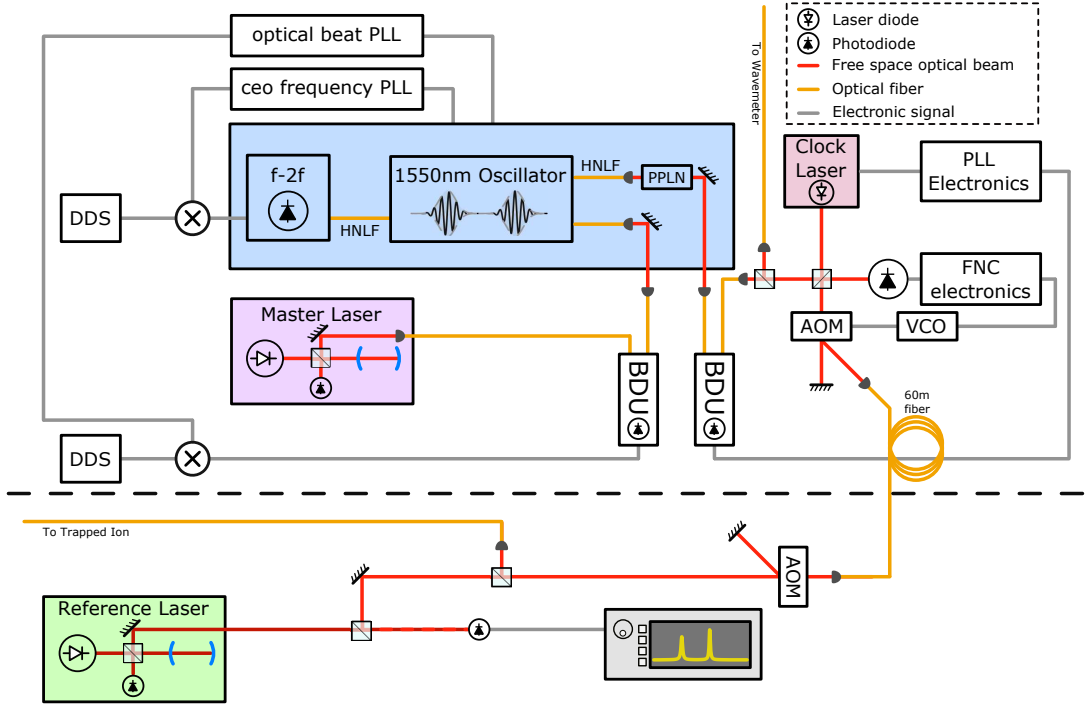} 
	\caption{Experimental setup scheme. The master laser is a \SI{1560}{nm} laser stabilized to a high finesse ULE cavity using PDH technique. The two comb RFs, $f_{ceo}$ and $f_{rep}$, are extracted by using a f-2f interferometer and an optical heterodyne beat of the master laser and one of the comb tooth, respectively. The clock laser is a 674nm ECDL which is fed into the first port of another heterodyne detector, with the the port being fed by a portion of the comb light that was coherently converted to have peak power about 674nm. As the comb inherits its stability from the master laser, the heterodyne signal is used to stabilize the clock laser, hence transfering the stability of the master laser to the clock laser}
	\label{fig: experimental scheme}
\end{figure*}
Our experimental setup is outlined in Fig. \ref{fig: experimental scheme}, including the phase-transfer scheme and the set-up used for comparisons of the comb-stabilized laser at \SI{674}{\nano\meter} with a cavity-stabilized laser operating at the same wavelength (hereafter referred to as the reference laser), and with the clock transition of a single trapped \textsuperscript{88}Sr\textsuperscript{+} ion. 

The master laser, a commercial laser source at \SI{1560}{\nano\meter} is Pound-Drever-Hall locked to an Ultra-Low Expansion (ULE) glass, high finesse, Fabry-Perot cavity, with servo bandwidth of approximately \SI{200}{\kilo\hertz}. The comb, an erbium-doped fiber mode-locked laser oscillator, outputs a comb spectrum centered around \SI{1560}{\nano\meter}, which is then amplified and split into several output ports. The carrier-envelope offset frequency, $f_{ceo}$, is locked using an f-2f interferometer \cite{Ye2005}. Small fractions ($\approx$1mW) of the master laser and comb main port outputs are injected into a fiber-coupled commercial beat detection unit (BDU) through optical fibers. We employ a PLL scheme to lock the BDU output, which is the beat signal of the master laser and the closest comb tooth, to a \SI{10}{\mega\hertz} GPS referenced DDS signal. The feedback loop controls $f_{r}$ with a \SI{200}{\kilo\hertz} servo bandwidth, transferring the phase stability of the master laser to the comb. 

The \SI{674}{nm} clock laser is an External-Cavity Diode Laser (ECDL). The laser output power is split into two paths, the first is coupled to a \SI{5}{\meter} long fiber connected to another BDU, while the second is coupled into a \SI{60}{\meter} fiber through a fiber noise-cancellation system. This long fiber leads to another lab where the comparison with the reference laser and the ion spectroscopy measurements have been carried out. 

Comb light going into a third optical output port is fed into a nonlinear optical fiber, increasing the spectral width of the pulse so that a large portion of the optical power is spread around \SI{1248}{nm}. It is then frequency doubled to produce a phase coherent comb spectrum at \SI{674}{nm}, which is coupled using a short fiber into the other port of that second BDU. The output beat at \SI{60}{\mega\hertz} is band-pass filtered, frequency divided and compared to a GPS referenced \SI{10}{\mega\hertz} oscillating signal. Using a PLL scheme with a digital frequency phase detector we then control a fast current modulation port of the \SI{674}{nm} laser diode to lock the two lasers beat, effectively reducing the laser phase noise while transferring the phase stability of the master laser to the clock laser. 

\begin{figure}
  	\centering
  		\begin{subfigure}[b]{\columnwidth}
    	\begin{minipage}[l]{0.01\columnwidth}
    		\caption{}
    		\label{fig: out of loop phase noise}
  		\end{minipage}
  		\hfil
    	\begin{minipage}[c]{0.85\columnwidth}
        	\includegraphics[width=\textwidth]{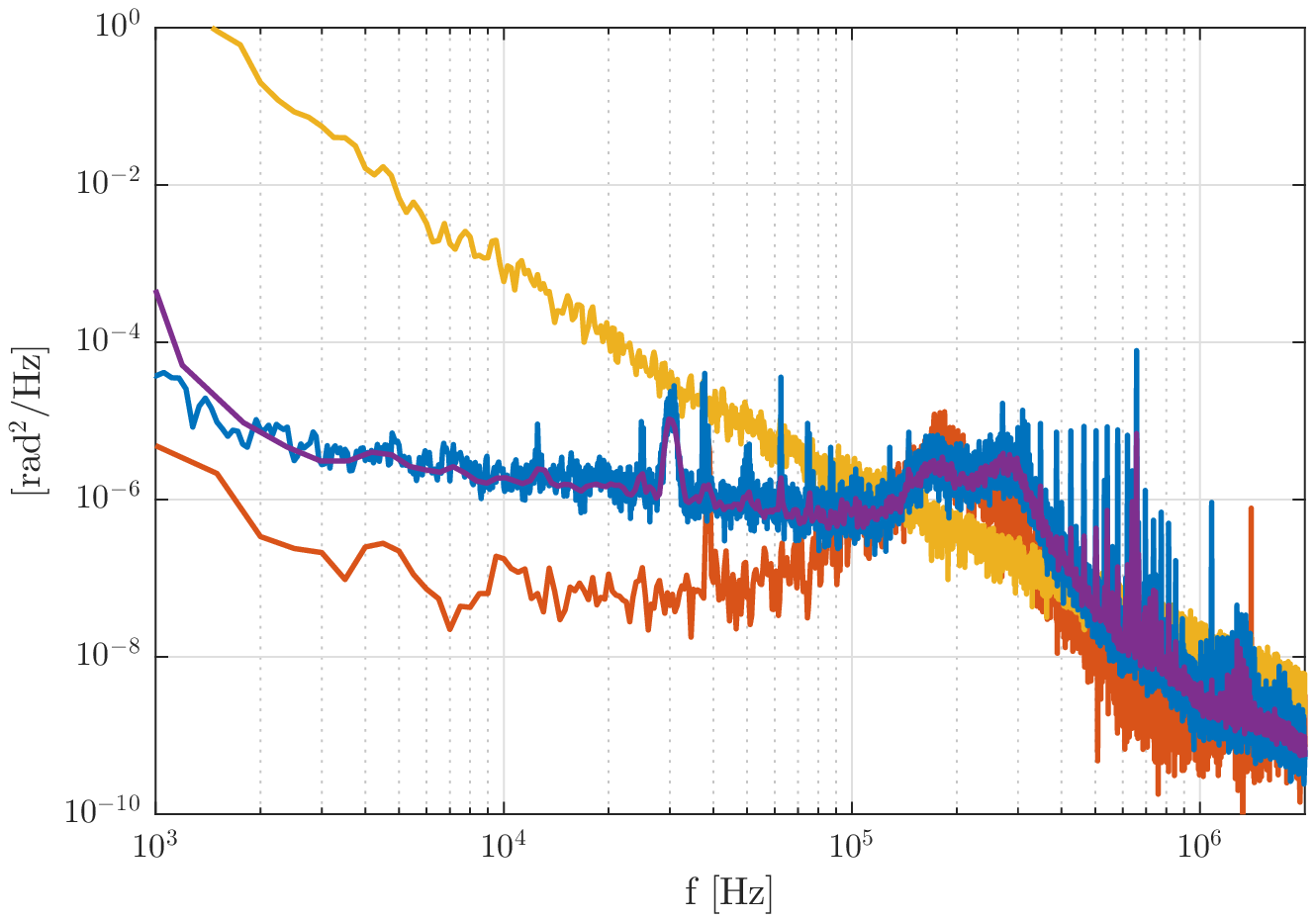}
  		\end{minipage}
    \end{subfigure}\\
  	\begin{subfigure}[b]{\columnwidth} 
  		\begin{minipage}[l]{0.01\columnwidth}
			\caption{}
    		\label{fig: out of loop psd}
  		\end{minipage}
  		\hfil
    	\begin{minipage}[c]{0.85\columnwidth}
    		\includegraphics[width=\columnwidth]{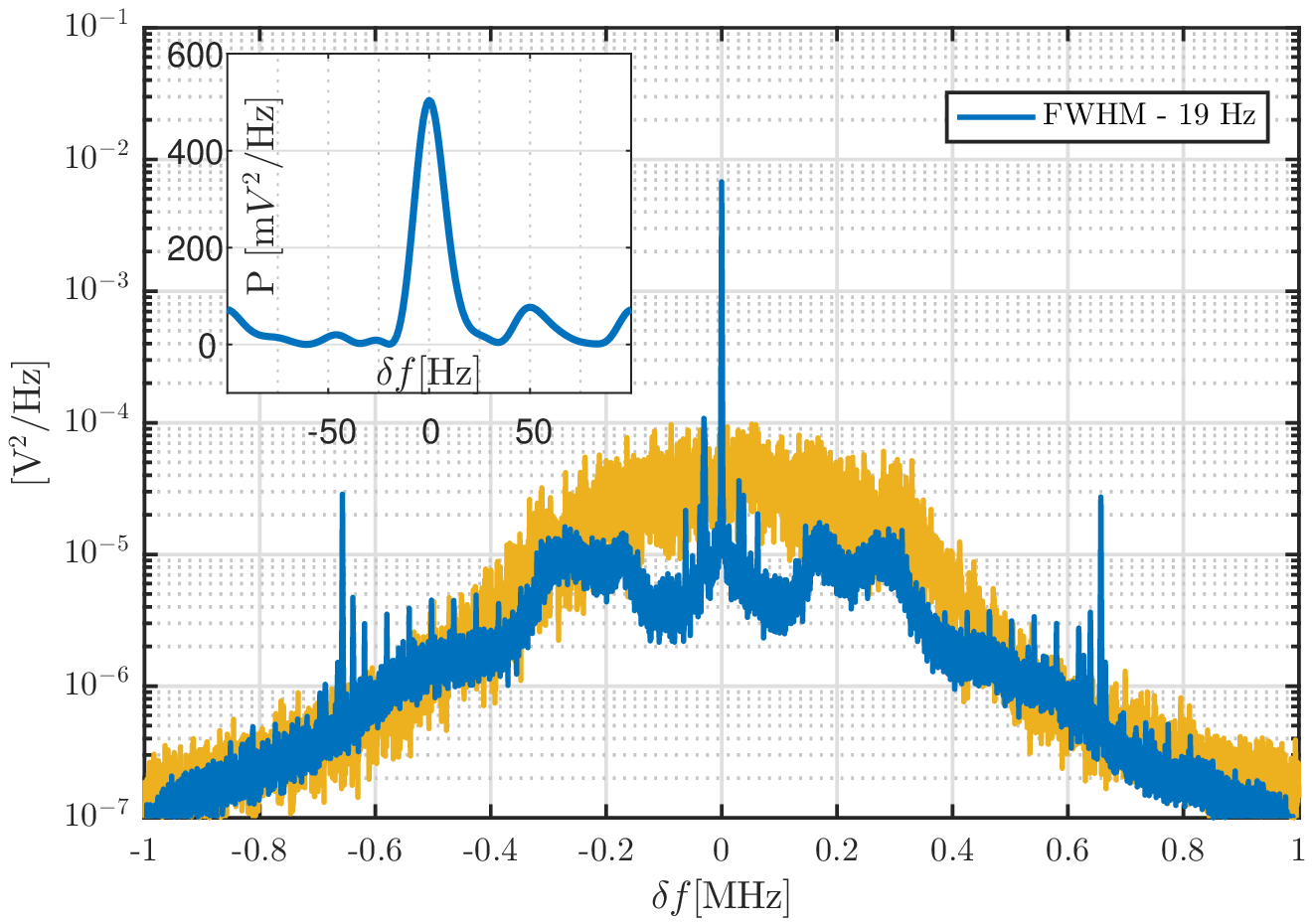}
  		\end{minipage}   
	\end{subfigure}%
  	\caption{Estimated power spectral densities of clock laser and frequency comb. Fig (a) present the phase noise PSD estimates for the comb tooth (red), free-running clock laser ( yellow) and locked clock laser (blue). The reduction of clock laser phase noise and the effect of the comb servo bumps on the clock laser spectrum can be appreciated. The purple spectrum is the spectrum of a synthesized phase noise process we used to simulate the ion-laser interaction. Fig (b) presents the PSD for the oscillating beat between the clock laser and the cavity stabilized laser, with the clock laser being unlocked (yellow) and locked (blue) to the frequency comb. The \SI{0.5}{\mega\hertz} lock bandwidth can be seen, as well as the large sidebands at \SI{200}{\kilo\hertz}. }
  	\label{fig: psds}
\end{figure}

To characterize the performance of our locking scheme we performed frequency beat measurements between the comb and clock lasers with the reference laser. In each case we recorded the oscillating beat signal and performed all other data processing on it. We extracted the phase noise signal from the time-dependent oscillating signal by implementing an all-digital phase noise measurement (detailed in \cite{angrisani2001digital}). We then used the phase noise signal to estimate the underlying power spectral density (PSD). In such way we characterized the phase noise of the comb spectral tooth at 674nm, the free running clock laser and locked clock laser. These measurements are presented in Fig. \ref{fig: psds}. 

We first inspected the phase noise signal of the comb tooth at \SI{674}{\nano\meter}, extracted from the beat with the reference laser. Since the reference laser is known to have a fast linewidth narrower than \SI{100}{\hertz}, we could attribute all fast phase noise in the beat signal to the frequency comb, hereby characterizing its fast phase modulations. The red solid line in Fig. \ref{fig: out of loop phase noise} shows the phase-noise PSD. A dominant feature it exhibits is the large increase in phase noise around \SI{200}{\kilo\hertz}. This large increase is the result of two contributions: The first contribution is that of the servo-bump of the control loop locking the comb repetition rate to the master laser, and the second is the servo bump of the master laser lock to the ULE cavity. The master laser servo bumps were independently observed in a self-heterodyne measurement of the master laser. The contribution of the repetition rate servo bumps was observed in the optical comb lock error-signal. We estimate the RMS phase noise, integrated over \SI{1}{\kilo \hertz} - \SI{2}{\mega\hertz}, to be 0.7 radian. \\
Next, we recorded the beat signal of the reference and clock lasers. We took two measurements; One in which the clock laser was free running and another in which it was locked to the comb tooth. The yellow (blue) solid line in Figure \ref{fig: out of loop psd} is the PSD of the free running (locked) beat signal. It is clear that when the clock laser is locked, a substantial amount of the signal energy is concentrated around the central peak, here evaluated to be below the \SI{1}{\kilo\hertz} resolution limit, whereas for the free running signal the energy is spread over a larger band. The inset in Fig. \ref{fig: out of loop psd} shows the same PSD, filtered with a \SI{30}{\kilo\hertz} band-pass filter and integrated over a time corresponding to a Fourier limit of \SI{12.5}{\hertz}. The full-width-half-max of the PSD signal is \SI{19}{\hertz}, placing a lower bound on the relative stability of the reference and the clock lasers.\\
To appreciate the spectral properties of the residual phase noise we inspected the PSD of the extracted phase noise signal. The yellow (blue) solid-line in Fig \ref{fig: out of loop phase noise} is the extracted phase noise PSD of the free-running (locked) clock laser and reference laser beat. The lock bandwidth is estimated to be roughly \SI{0.5}{\MHz}, limited by the clock laser modulation response. The master laser and comb servo bumps at around \SI{200}{\kilo\hertz} are therefore well within the loop bandwidth and are hence transferred to the clock laser from the frequency comb. Other apparent features are  frequency spikes in the $10^4-10^5$\si{\hertz} and the $3\cdot10^5-10^6$\si{\hertz} frequency bands, both are likely due to electric noise in the lab ground signal, which we where unable to reduce. We measured the RMS phase noise of the locked clock laser, integrated over the same band as above, to be 2.2 radians; i.e. roughly three times larger than that of the comb tooth.\\

To further test the spectral properties of our clock laser we used it to drive the $\ket{4S_{1/2}}\to\ket{4D_{5/2}}$ electric-quadruple optical clock transition of a single \textsuperscript{88}Sr\textsuperscript{+} ion, trapped in a linear Paul trap. Details of the Paul-trap apparatus and the lasers involved in trapping, cooling and imaging the ion can be found in \cite{Akerman2012}. Optical clock transition spectroscopy serves to characterize the frequency noise spectrum between the clock laser and the electronic transition. 

In a first experiment we performed Rabi nutation on the clock transition by tuning the laser to resonance and, at fixed laser power, scanning the pulse duration. We initiated our ion in the $\ket{4S_{1/2}}$ state, and performed 100 repetitions of each pulse time. We then used a maximum likelihood fit to extract the decay of Rabi oscillations. Rabi oscillation amplitude can decay due to different noise sources, including the laser-ion phase noise we measured in the beat measurements described above.

\begin{figure}[t]
  	\centering
  	\includegraphics[width=\columnwidth]{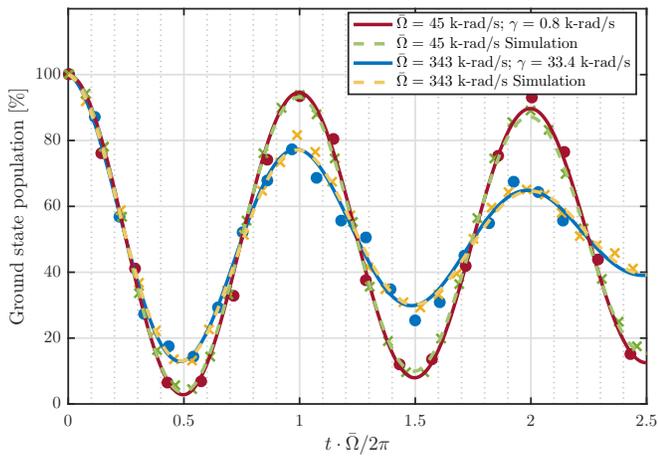}
    \caption{Rabi oscillations over normalized time. Oscillations of the ground state population where observed when driving the two levels both with high (blue trend and dots) and low (red trend and dots) laser power, corresponding to different Rabi frequencies. The de-coherence rate measured was significantly higher when high laser power was used. The yellow and green trends are simulation results of the interaction of a similarly noisy laser and ion, verifying the relation of the decoherence to the laser phase noise spectrum}
    \label{fig: rabi oscillations}
\end{figure}

The data (solid dots) and fit (solid lines) of two such measurements are shown in Fig. \ref{fig: rabi oscillations}. The two measurements differ by the clock laser optical power, corresponding to different Rabi frequencies; 45 kHz and 343 kHZ respectively. For ease of comparison the two nutation curves are shown on a time-axis which is normalized by the Rabi nutation period. As seen, the slower Rabi oscillations (red curve) maintain better coherence whereas fast Rabi oscillations (blue curves) de-cohere after $\sim$3 Rabi periods. This is a manifestation of the high-frequency laser phase noise inherited from the frequency comb. 

To test whether this assumption is correct we performed a numerical simulation of the Rabi nutation experiment. We numerically generated a phase noise signal with the same PSD as was measured by the beat with the reference laser (purple solid line in Fig. \ref{fig: out of loop phase noise}), and used it to realize a stochastic laser phase term in a monte-carlo simulation of two level system interacting with laser. The simulation results are shown by the light yellow and green crosses and dashed lines in Fig. \ref{fig: rabi oscillations}. As seen, the laser phase noise is indeed the main source of decoherence. A similar effect was observed in \cite{Scharnhorst2015}, where the authors have implemented a transfer oscillator scheme which mitigated the frequency comb induced phase noise and resulted in increased coherence for fast Rabi nutations.\\

We next turned to examine the ion-laser coherence at longer times by performing Rabi frequency scans. Here we used a \SI{8.25}{\milli\second} $\pi$-pulse corresponding to a \SI{107}{\hertz} Fourier limited spectral width.  
\begin{figure}[t]
  	\centering
  	\includegraphics[width=\columnwidth]{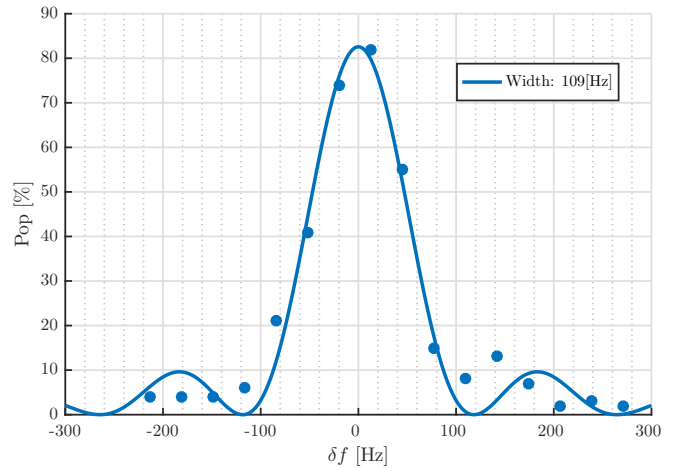}
    \caption{Rabi spectroscopy with long pulse time. The measured pulse width, \SI{109}{\hertz}, corresponds to the pulse Fourier limit, which is \SI{107}{\hertz}}
    \label{fig: rabi spectroscopy}
\end{figure}
The results of a laser frequency scan across the ion resonance frequency are shown in Fig. \ref{fig: rabi spectroscopy}. The FWHM of the Rabi resonance is \SI{109}{\hertz} ($\pm$\SI{14}{\hertz}) consistent with finite-time Fourier broadening. Longer pulses did not result further narrowing of the resonance indicating that at times longer than \SI{8.25}{\milli\second}, we are limited by low frequency phase noise of the relative frequency between the laser and the ion.\\ 

To investigate whether this relative phase noise is dominated by the laser or the ion we converted to a different spectroscopic scheme. A dominant source of noise in the $\ket{S}\to\ket{D}$  transition frequency is due to ambient magnetic noise via the first order Zeeman effect. To mitigate the resulting phase noise we performed a Ramsey spectroscopy experiment on the optical clock transition, incorporating a series of radio-frequency Magnetic Field Dynamical Decoupling (MFDD) pulses during the Ramsey wait time \cite{Akerman2015}. The magnetic susceptibility of the $\ket{S}\to\ket{D}$ transition frequency is,
\begin{align}
 \begin{split} \label{eq: magnetic sus}
 \chi(\ket{S,m_S}&\to \ket{D,m_D}) =\\ &-2.802\cdot m_S + 1.68\cdot m_D\quad \text{[MHz/G]},
 \end{split}
 \end{align} 
where $m_S$ and $m_D$ are the respective Zeeman sub-levels. The MFDD pulses coherently transfer the superposition in the $S$ or the $D$ manifold, between two chosen $m$ levels, such that the resulting susceptibilities in the two cases have opposite signs. The ratio of time intervals between subsequent MFDD pulses is chosen to match the ratio of susceptibilities of the two $m$ levels. Thus the accumulated phase due to (low frequency) magnetic field noise is reversed in each two-pulse cycle. 
 
 \begin{figure}
  	\centering
  	\begin{subfigure}[b]{\columnwidth} 
  		\begin{minipage}[l]{0.01\columnwidth}
			\caption{}
    		\label{fig: mfdd scheme}
  		\end{minipage}
  		\hfil
    	\begin{minipage}[c]{0.85\columnwidth}
    		\includegraphics[width=\columnwidth]{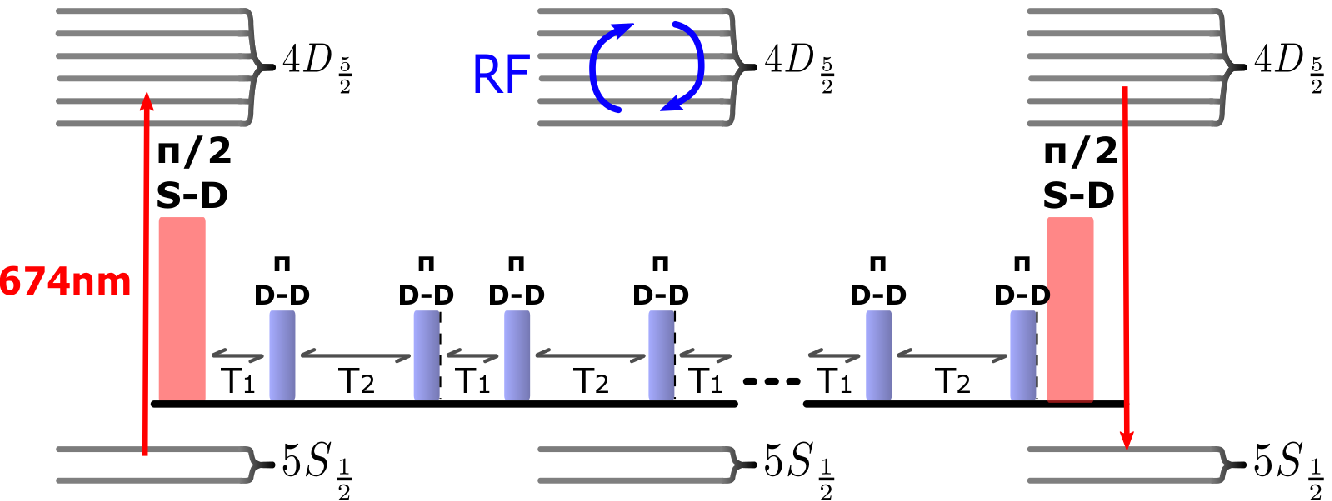}
  		\end{minipage}   
	\end{subfigure}\\
  	\begin{subfigure}[b]{\columnwidth}
    	\begin{minipage}[l]{0.01\columnwidth}
    		\caption{}
    		\label{fig: ramseys}
  		\end{minipage}
  		\hfil
    	\begin{minipage}[c]{0.85\columnwidth}
        	\includegraphics[width=\textwidth]{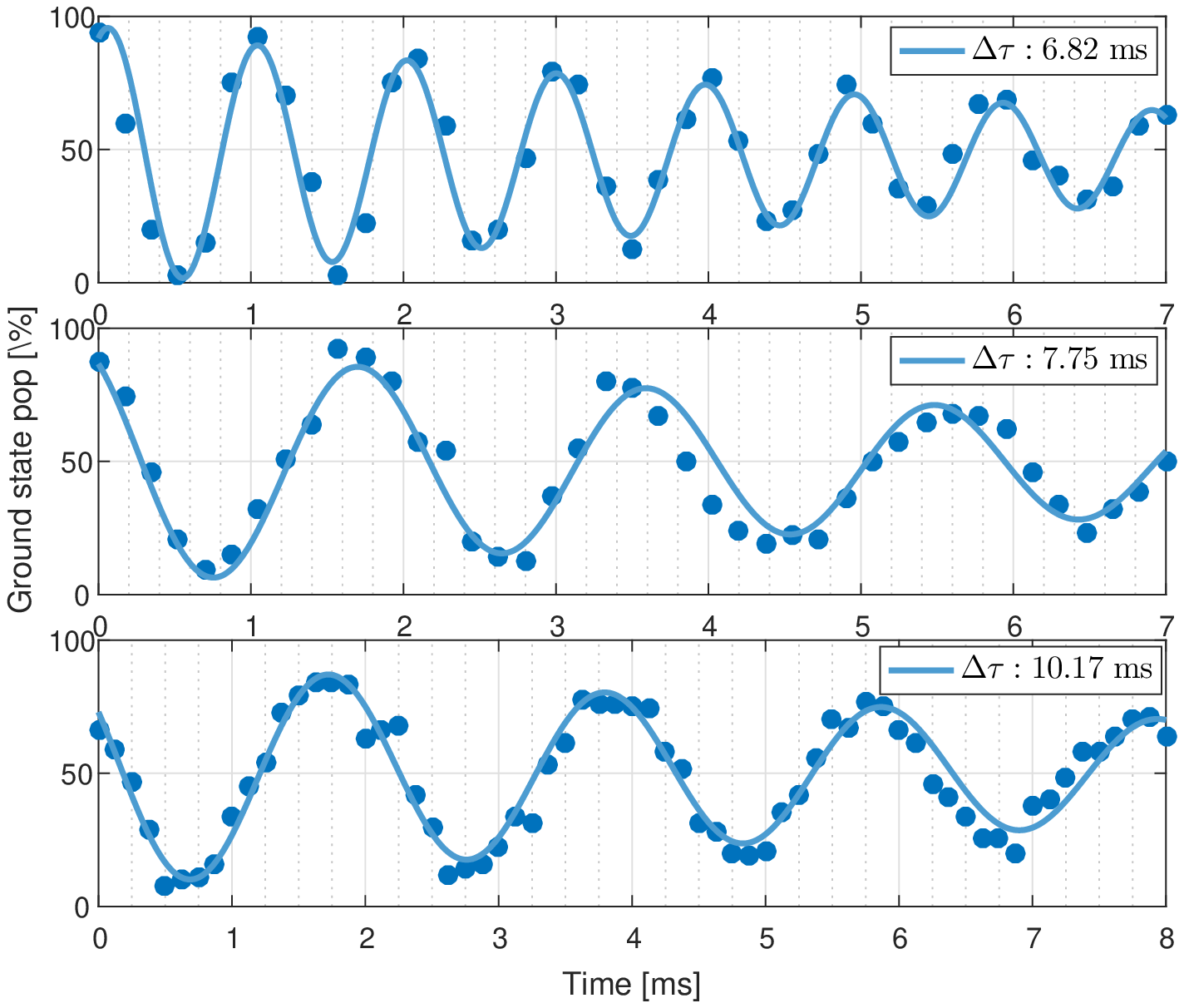}
  		\end{minipage}
    \end{subfigure}%
  	\caption{Ramsey spectroscopy with MFDD pulses. Figure (a) schematically depicts the experimental sequence, starting with a $\pi/2$ pulse on the optical transition, followed by a series of RF $\pi$-pulses flipping the population between the $m_D = \pm 3/2$ states, with the time between pulses $T_1$ and $T_2$ determined by the magnetic susceptibility ratio. The series is closed with a $\pi /2$ pulse and the ion is measured. Figure (b) shows the results for such measurement, for no MFDD Pulse (top graph), one MFDD pulse (middle graph) and four MFDD pulses (bottom graph). The increased phase coherence, manifested by the slower de-coherence rate, proves that the laser-ion combined phase noise was dominated by magnetic field noise}
  	\label{fig: ramsey mfdd}
\end{figure} 

We performed a series of Ramsey spectroscopy experiments with different number of MFDD pulses. Figure \ref{fig: ramsey mfdd} shows both a schematic time-sequence and the results for three such experiments, using zero, one and four MFDD pulses. We started with driving the ion on the $\ket{S,-1/2}\leftrightarrow\ket{D,-3/2}$ transition, preparing an equal superposition of these states. The RF pulses where chosen to drive the $\ket{D,-3/2}\leftrightarrow\ket{D,3/2}$ transition in the $4D_{3/2}$ manifold. The time between pulses $T_1$ was determined according to,
\begin{align}
\begin{split}
T_1 = \frac{T}{N(1+r)},\quad T_2 =  T_1\cdot r.
\end{split}
\end{align}
Here, N is number of MFDD pulses and r is the susceptibility ratio according to Eq. (\ref{eq: magnetic sus}). As seen, the three measurements show an increase in coherence time as the number of MFDD pulses increases. This is an indication for a significant contribution of magnetic field noise to the overall ion-laser phase noise comparison. Adding more than three pulses did not prolong our coherence time further, implying that the remaining phase noise is not due to magnetic field noise. Using four MFDD pulses we measured a coherence time of $\Delta\tau =$\SI{10.17}{\milli\second}. Assuming a white frequency noise model, this corresponds to a laser linewidth of \SI{16}{\hertz}\cite{agarwal1978quantum}, relatively close to the \SI{19}{\hertz} obtained by beating the the clock laser and the reference laser.\\ 
\section{conclusion}

In this work we demonstrated the transfer of phase stability of a stable, reference at telecom wavelength across the optical spectrum to visible light at \SI{674}{\nano\meter}. This scheme is especially useful when building systems in which the spectroscopy of multiple species and transitions requires narrow-linewidth lasers of multiple wavelengths. Here, the phase stability of a single reference can be conveniently distributed across optical spectral gas to multiple wavelengths.  We measured the linewidth of our phase-stable laser at 674 nm by beating it against another, independently stabilized, narrow-linewidth laser and by performing precision spectroscopy of a single trapped-ion. Both measurements exhibited the linewidth reduction of the bare laser diode, with the two laser comparison indicating a linewidth \SI{19}{\hertz} after \SI{8.25}{\milli\second} and the spectroscopic measurement a linewidth of \SI{16}{\hertz} after \SI{8}{ms}. On the other hand, the fast locking of the laser to the comb transferred the fast phase noise component resulting from the servo bumps of the comb to the master laser and the master laser to the reference cavity. We have shown that this high-frequency phase-noise imposes decoherence when performing fast operations on the ion. 

This work was supported by the Crown Photonics Center, ICore-Israeli excellence center
circle of light, the Israeli Science Foundation, the Israeli Ministry of Science Technology
and Space, the Minerva Stiftung and the European Research Council (consolidator grant 616919-Ionology).

\bibliographystyle{apsrev4-1} 
\bibliography{phase_stability_transfer} 

\begin{thebibliography}{35}%
\makeatletter
\providecommand \@ifxundefined [1]{%
 \@ifx{#1\undefined}
}%
\providecommand \@ifnum [1]{%
 \ifnum #1\expandafter \@firstoftwo
 \else \expandafter \@secondoftwo
 \fi
}%
\providecommand \@ifx [1]{%
 \ifx #1\expandafter \@firstoftwo
 \else \expandafter \@secondoftwo
 \fi
}%
\providecommand \natexlab [1]{#1}%
\providecommand \enquote  [1]{``#1''}%
\providecommand \bibnamefont  [1]{#1}%
\providecommand \bibfnamefont [1]{#1}%
\providecommand \citenamefont [1]{#1}%
\providecommand \href@noop [0]{\@secondoftwo}%
\providecommand \href [0]{\begingroup \@sanitize@url \@href}%
\providecommand \@href[1]{\@@startlink{#1}\@@href}%
\providecommand \@@href[1]{\endgroup#1\@@endlink}%
\providecommand \@sanitize@url [0]{\catcode `\\12\catcode `\$12\catcode
  `\&12\catcode `\#12\catcode `\^12\catcode `\_12\catcode `\%12\relax}%
\providecommand \@@startlink[1]{}%
\providecommand \@@endlink[0]{}%
\providecommand \url  [0]{\begingroup\@sanitize@url \@url }%
\providecommand \@url [1]{\endgroup\@href {#1}{\urlprefix }}%
\providecommand \urlprefix  [0]{URL }%
\providecommand \Eprint [0]{\href }%
\providecommand \doibase [0]{http://dx.doi.org/}%
\providecommand \selectlanguage [0]{\@gobble}%
\providecommand \bibinfo  [0]{\@secondoftwo}%
\providecommand \bibfield  [0]{\@secondoftwo}%
\providecommand \translation [1]{[#1]}%
\providecommand \BibitemOpen [0]{}%
\providecommand \bibitemStop [0]{}%
\providecommand \bibitemNoStop [0]{.\EOS\space}%
\providecommand \EOS [0]{\spacefactor3000\relax}%
\providecommand \BibitemShut  [1]{\csname bibitem#1\endcsname}%
\let\auto@bib@innerbib\@empty
\bibitem [{\citenamefont {Ludlow}\ \emph {et~al.}(2015)\citenamefont {Ludlow},
  \citenamefont {Boyd}, \citenamefont {Ye}, \citenamefont {Peik},\ and\
  \citenamefont {Schmidt}}]{ludlow2015optical}%
  \BibitemOpen
  \bibfield  {author} {\bibinfo {author} {\bibfnamefont {A.~D.}\ \bibnamefont
  {Ludlow}}, \bibinfo {author} {\bibfnamefont {M.~M.}\ \bibnamefont {Boyd}},
  \bibinfo {author} {\bibfnamefont {J.}~\bibnamefont {Ye}}, \bibinfo {author}
  {\bibfnamefont {E.}~\bibnamefont {Peik}}, \ and\ \bibinfo {author}
  {\bibfnamefont {P.~O.}\ \bibnamefont {Schmidt}},\ }\href@noop {} {\bibfield
  {journal} {\bibinfo  {journal} {Reviews of Modern Physics}\ }\textbf
  {\bibinfo {volume} {87}},\ \bibinfo {pages} {637} (\bibinfo {year}
  {2015})}\BibitemShut {NoStop}%
\bibitem [{\citenamefont {Nicholson}\ \emph {et~al.}(2012)\citenamefont
  {Nicholson}, \citenamefont {Martin}, \citenamefont {Williams}, \citenamefont
  {Bloom}, \citenamefont {Bishof}, \citenamefont {Swallows}, \citenamefont
  {Campbell},\ and\ \citenamefont {Ye}}]{nicholson2012comparison}%
  \BibitemOpen
  \bibfield  {author} {\bibinfo {author} {\bibfnamefont {T.}~\bibnamefont
  {Nicholson}}, \bibinfo {author} {\bibfnamefont {M.}~\bibnamefont {Martin}},
  \bibinfo {author} {\bibfnamefont {J.}~\bibnamefont {Williams}}, \bibinfo
  {author} {\bibfnamefont {B.}~\bibnamefont {Bloom}}, \bibinfo {author}
  {\bibfnamefont {M.}~\bibnamefont {Bishof}}, \bibinfo {author} {\bibfnamefont
  {M.}~\bibnamefont {Swallows}}, \bibinfo {author} {\bibfnamefont
  {S.}~\bibnamefont {Campbell}}, \ and\ \bibinfo {author} {\bibfnamefont
  {J.}~\bibnamefont {Ye}},\ }\href@noop {} {\bibfield  {journal} {\bibinfo
  {journal} {Physical review letters}\ }\textbf {\bibinfo {volume} {109}},\
  \bibinfo {pages} {230801} (\bibinfo {year} {2012})}\BibitemShut {NoStop}%
\bibitem [{\citenamefont {Hinkley}\ \emph {et~al.}(2013)\citenamefont
  {Hinkley}, \citenamefont {Sherman}, \citenamefont {Phillips}, \citenamefont
  {Schioppo}, \citenamefont {Lemke}, \citenamefont {Beloy}, \citenamefont
  {Pizzocaro}, \citenamefont {Oates},\ and\ \citenamefont
  {Ludlow}}]{hinkley2013atomic}%
  \BibitemOpen
  \bibfield  {author} {\bibinfo {author} {\bibfnamefont {N.}~\bibnamefont
  {Hinkley}}, \bibinfo {author} {\bibfnamefont {J.}~\bibnamefont {Sherman}},
  \bibinfo {author} {\bibfnamefont {N.}~\bibnamefont {Phillips}}, \bibinfo
  {author} {\bibfnamefont {M.}~\bibnamefont {Schioppo}}, \bibinfo {author}
  {\bibfnamefont {N.}~\bibnamefont {Lemke}}, \bibinfo {author} {\bibfnamefont
  {K.}~\bibnamefont {Beloy}}, \bibinfo {author} {\bibfnamefont
  {M.}~\bibnamefont {Pizzocaro}}, \bibinfo {author} {\bibfnamefont {C.~W.}\
  \bibnamefont {Oates}}, \ and\ \bibinfo {author} {\bibfnamefont
  {A.}~\bibnamefont {Ludlow}},\ }\href@noop {} {\bibfield  {journal} {\bibinfo
  {journal} {Science}\ }\textbf {\bibinfo {volume} {341}},\ \bibinfo {pages}
  {1215} (\bibinfo {year} {2013})}\BibitemShut {NoStop}%
\bibitem [{\citenamefont {Chou}\ \emph {et~al.}(2010)\citenamefont {Chou},
  \citenamefont {Hume}, \citenamefont {Rosenband},\ and\ \citenamefont
  {Wineland}}]{chou2010optical}%
  \BibitemOpen
  \bibfield  {author} {\bibinfo {author} {\bibfnamefont {C.-W.}\ \bibnamefont
  {Chou}}, \bibinfo {author} {\bibfnamefont {D.}~\bibnamefont {Hume}}, \bibinfo
  {author} {\bibfnamefont {T.}~\bibnamefont {Rosenband}}, \ and\ \bibinfo
  {author} {\bibfnamefont {D.}~\bibnamefont {Wineland}},\ }\href@noop {}
  {\bibfield  {journal} {\bibinfo  {journal} {Science}\ }\textbf {\bibinfo
  {volume} {329}},\ \bibinfo {pages} {1630} (\bibinfo {year}
  {2010})}\BibitemShut {NoStop}%
\bibitem [{\citenamefont {Eisele}\ \emph {et~al.}(2009)\citenamefont {Eisele},
  \citenamefont {Nevsky},\ and\ \citenamefont
  {Schiller}}]{eisele2009laboratory}%
  \BibitemOpen
  \bibfield  {author} {\bibinfo {author} {\bibfnamefont {C.}~\bibnamefont
  {Eisele}}, \bibinfo {author} {\bibfnamefont {A.~Y.}\ \bibnamefont {Nevsky}},
  \ and\ \bibinfo {author} {\bibfnamefont {S.}~\bibnamefont {Schiller}},\
  }\href@noop {} {\bibfield  {journal} {\bibinfo  {journal} {Physical Review
  Letters}\ }\textbf {\bibinfo {volume} {103}},\ \bibinfo {pages} {090401}
  (\bibinfo {year} {2009})}\BibitemShut {NoStop}%
\bibitem [{\citenamefont {Abbott}\ \emph {et~al.}(2016)\citenamefont {Abbott},
  \citenamefont {Abbott}, \citenamefont {Abbott}, \citenamefont {Abernathy},
  \citenamefont {Acernese}, \citenamefont {Ackley}, \citenamefont {Adams},
  \citenamefont {Adams}, \citenamefont {Addesso}, \citenamefont {Adhikari}
  \emph {et~al.}}]{abbott2016observation}%
  \BibitemOpen
  \bibfield  {author} {\bibinfo {author} {\bibfnamefont {B.~P.}\ \bibnamefont
  {Abbott}}, \bibinfo {author} {\bibfnamefont {R.}~\bibnamefont {Abbott}},
  \bibinfo {author} {\bibfnamefont {T.}~\bibnamefont {Abbott}}, \bibinfo
  {author} {\bibfnamefont {M.}~\bibnamefont {Abernathy}}, \bibinfo {author}
  {\bibfnamefont {F.}~\bibnamefont {Acernese}}, \bibinfo {author}
  {\bibfnamefont {K.}~\bibnamefont {Ackley}}, \bibinfo {author} {\bibfnamefont
  {C.}~\bibnamefont {Adams}}, \bibinfo {author} {\bibfnamefont
  {T.}~\bibnamefont {Adams}}, \bibinfo {author} {\bibfnamefont
  {P.}~\bibnamefont {Addesso}}, \bibinfo {author} {\bibfnamefont
  {R.}~\bibnamefont {Adhikari}},  \emph {et~al.},\ }\href@noop {} {\bibfield
  {journal} {\bibinfo  {journal} {Physical review letters}\ }\textbf {\bibinfo
  {volume} {116}},\ \bibinfo {pages} {061102} (\bibinfo {year}
  {2016})}\BibitemShut {NoStop}%
\bibitem [{\citenamefont {Drever}\ \emph {et~al.}(1983)\citenamefont {Drever},
  \citenamefont {Hall}, \citenamefont {Kowalski}, \citenamefont {Hough},
  \citenamefont {Ford}, \citenamefont {Munley},\ and\ \citenamefont
  {Ward}}]{Drever1983}%
  \BibitemOpen
  \bibfield  {author} {\bibinfo {author} {\bibfnamefont {R.~W.~P.}\
  \bibnamefont {Drever}}, \bibinfo {author} {\bibfnamefont {J.~L.}\
  \bibnamefont {Hall}}, \bibinfo {author} {\bibfnamefont {F.~V.}\ \bibnamefont
  {Kowalski}}, \bibinfo {author} {\bibfnamefont {J.}~\bibnamefont {Hough}},
  \bibinfo {author} {\bibfnamefont {G.~M.}\ \bibnamefont {Ford}}, \bibinfo
  {author} {\bibfnamefont {A.~J.}\ \bibnamefont {Munley}}, \ and\ \bibinfo
  {author} {\bibfnamefont {H.}~\bibnamefont {Ward}},\ }\href {\doibase
  10.1007/BF00702605} {\bibfield  {journal} {\bibinfo  {journal} {Applied
  Physics B}\ }\textbf {\bibinfo {volume} {31}},\ \bibinfo {pages} {97}
  (\bibinfo {year} {1983})}\BibitemShut {NoStop}%
\bibitem [{\citenamefont {Swallows}\ \emph {et~al.}(2012)\citenamefont
  {Swallows}, \citenamefont {Martin}, \citenamefont {Bishof}, \citenamefont
  {Benko}, \citenamefont {Lin}, \citenamefont {Blatt}, \citenamefont {Rey},\
  and\ \citenamefont {Ye}}]{Swallows2012}%
  \BibitemOpen
  \bibfield  {author} {\bibinfo {author} {\bibfnamefont {M.}~\bibnamefont
  {Swallows}}, \bibinfo {author} {\bibfnamefont {M.}~\bibnamefont {Martin}},
  \bibinfo {author} {\bibfnamefont {M.}~\bibnamefont {Bishof}}, \bibinfo
  {author} {\bibfnamefont {C.}~\bibnamefont {Benko}}, \bibinfo {author}
  {\bibfnamefont {Y.}~\bibnamefont {Lin}}, \bibinfo {author} {\bibfnamefont
  {S.}~\bibnamefont {Blatt}}, \bibinfo {author} {\bibfnamefont {A.~M.}\
  \bibnamefont {Rey}}, \ and\ \bibinfo {author} {\bibfnamefont
  {J.}~\bibnamefont {Ye}},\ }\href {\doibase 10.1109/TUFFC.2012.2210}
  {\bibfield  {journal} {\bibinfo  {journal} {IEEE transactions on ultrasonics,
  ferroelectrics, and frequency control}\ }\textbf {\bibinfo {volume} {59}},\
  \bibinfo {pages} {416} (\bibinfo {year} {2012})}\BibitemShut {NoStop}%
\bibitem [{\citenamefont {Alnis}\ \emph {et~al.}(2008)\citenamefont {Alnis},
  \citenamefont {Matveev}, \citenamefont {Kolachevsky}, \citenamefont {Udem},\
  and\ \citenamefont {H\"ansch}}]{Alnis2008}%
  \BibitemOpen
  \bibfield  {author} {\bibinfo {author} {\bibfnamefont {J.}~\bibnamefont
  {Alnis}}, \bibinfo {author} {\bibfnamefont {A.}~\bibnamefont {Matveev}},
  \bibinfo {author} {\bibfnamefont {N.}~\bibnamefont {Kolachevsky}}, \bibinfo
  {author} {\bibfnamefont {T.}~\bibnamefont {Udem}}, \ and\ \bibinfo {author}
  {\bibfnamefont {T.~W.}\ \bibnamefont {H\"ansch}},\ }\href {\doibase
  10.1103/PhysRevA.77.053809} {\bibfield  {journal} {\bibinfo  {journal} {Phys.
  Rev. A}\ }\textbf {\bibinfo {volume} {77}},\ \bibinfo {pages} {053809}
  (\bibinfo {year} {2008})}\BibitemShut {NoStop}%
\bibitem [{\citenamefont {Young}\ \emph {et~al.}(1999)\citenamefont {Young},
  \citenamefont {Cruz}, \citenamefont {Itano},\ and\ \citenamefont
  {Bergquist}}]{B.C.YoungW.M.Itano1999}%
  \BibitemOpen
  \bibfield  {author} {\bibinfo {author} {\bibfnamefont {B.~C.}\ \bibnamefont
  {Young}}, \bibinfo {author} {\bibfnamefont {F.~C.}\ \bibnamefont {Cruz}},
  \bibinfo {author} {\bibfnamefont {W.~M.}\ \bibnamefont {Itano}}, \ and\
  \bibinfo {author} {\bibfnamefont {J.~C.}\ \bibnamefont {Bergquist}},\ }\href
  {\doibase 10.1103/PhysRevLett.82.3799} {\bibfield  {journal} {\bibinfo
  {journal} {Phys. Rev. Lett.}\ }\textbf {\bibinfo {volume} {82}},\ \bibinfo
  {pages} {3799} (\bibinfo {year} {1999})}\BibitemShut {NoStop}%
\bibitem [{\citenamefont {Jiang}\ \emph {et~al.}(2011)\citenamefont {Jiang},
  \citenamefont {Ludlow}, \citenamefont {Lemke}, \citenamefont {Fox},
  \citenamefont {Sherman}, \citenamefont {Ma},\ and\ \citenamefont
  {Oates}}]{Jiang2011}%
  \BibitemOpen
  \bibfield  {author} {\bibinfo {author} {\bibfnamefont {Y.}~\bibnamefont
  {Jiang}}, \bibinfo {author} {\bibfnamefont {A.}~\bibnamefont {Ludlow}},
  \bibinfo {author} {\bibfnamefont {N.~D.}\ \bibnamefont {Lemke}}, \bibinfo
  {author} {\bibfnamefont {R.~W.}\ \bibnamefont {Fox}}, \bibinfo {author}
  {\bibfnamefont {J.~A.}\ \bibnamefont {Sherman}}, \bibinfo {author}
  {\bibfnamefont {L.-S.}\ \bibnamefont {Ma}}, \ and\ \bibinfo {author}
  {\bibfnamefont {C.~W.}\ \bibnamefont {Oates}},\ }\href@noop {} {\bibfield
  {journal} {\bibinfo  {journal} {Nature Photonics}\ }\textbf {\bibinfo
  {volume} {5}},\ \bibinfo {pages} {158} (\bibinfo {year} {2011})}\BibitemShut
  {NoStop}%
\bibitem [{\citenamefont {Kessler}\ \emph
  {et~al.}(2012{\natexlab{a}})\citenamefont {Kessler}, \citenamefont {Legero},\
  and\ \citenamefont {Sterr}}]{Kessler2011}%
  \BibitemOpen
  \bibfield  {author} {\bibinfo {author} {\bibfnamefont {T.}~\bibnamefont
  {Kessler}}, \bibinfo {author} {\bibfnamefont {T.}~\bibnamefont {Legero}}, \
  and\ \bibinfo {author} {\bibfnamefont {U.}~\bibnamefont {Sterr}},\ }\href
  {\doibase 10.1364/JOSAB.29.000178} {\bibfield  {journal} {\bibinfo  {journal}
  {Journal of the Optical Society of America B}\ }\textbf {\bibinfo {volume}
  {29}},\ \bibinfo {pages} {178} (\bibinfo {year}
  {2012}{\natexlab{a}})}\BibitemShut {NoStop}%
\bibitem [{\citenamefont {Kessler}\ \emph
  {et~al.}(2012{\natexlab{b}})\citenamefont {Kessler}, \citenamefont
  {Hagemann}, \citenamefont {Grebing}, \citenamefont {Legero}, \citenamefont
  {Sterr}, \citenamefont {Riehle}, \citenamefont {Martin}, \citenamefont
  {Chen},\ and\ \citenamefont {Ye}}]{Kessler2012}%
  \BibitemOpen
  \bibfield  {author} {\bibinfo {author} {\bibfnamefont {T.}~\bibnamefont
  {Kessler}}, \bibinfo {author} {\bibfnamefont {C.}~\bibnamefont {Hagemann}},
  \bibinfo {author} {\bibfnamefont {C.}~\bibnamefont {Grebing}}, \bibinfo
  {author} {\bibfnamefont {T.}~\bibnamefont {Legero}}, \bibinfo {author}
  {\bibfnamefont {U.}~\bibnamefont {Sterr}}, \bibinfo {author} {\bibfnamefont
  {F.}~\bibnamefont {Riehle}}, \bibinfo {author} {\bibfnamefont
  {M.}~\bibnamefont {Martin}}, \bibinfo {author} {\bibfnamefont
  {L.}~\bibnamefont {Chen}}, \ and\ \bibinfo {author} {\bibfnamefont
  {J.}~\bibnamefont {Ye}},\ }\href@noop {} {\bibfield  {journal} {\bibinfo
  {journal} {Nature Photonics}\ }\textbf {\bibinfo {volume} {6}},\ \bibinfo
  {pages} {687} (\bibinfo {year} {2012}{\natexlab{b}})}\BibitemShut {NoStop}%
\bibitem [{\citenamefont {Hagemann}\ \emph {et~al.}(2013)\citenamefont
  {Hagemann}, \citenamefont {Grebing}, \citenamefont {Kessler}, \citenamefont
  {Falke}, \citenamefont {Lemke}, \citenamefont {Lisdat}, \citenamefont
  {Schnatz}, \citenamefont {Riehle},\ and\ \citenamefont
  {Sterr}}]{Hagemann2013}%
  \BibitemOpen
  \bibfield  {author} {\bibinfo {author} {\bibfnamefont {C.}~\bibnamefont
  {Hagemann}}, \bibinfo {author} {\bibfnamefont {C.}~\bibnamefont {Grebing}},
  \bibinfo {author} {\bibfnamefont {T.}~\bibnamefont {Kessler}}, \bibinfo
  {author} {\bibfnamefont {S.}~\bibnamefont {Falke}}, \bibinfo {author}
  {\bibfnamefont {N.}~\bibnamefont {Lemke}}, \bibinfo {author} {\bibfnamefont
  {C.}~\bibnamefont {Lisdat}}, \bibinfo {author} {\bibfnamefont
  {H.}~\bibnamefont {Schnatz}}, \bibinfo {author} {\bibfnamefont
  {F.}~\bibnamefont {Riehle}}, \ and\ \bibinfo {author} {\bibfnamefont
  {U.}~\bibnamefont {Sterr}},\ }\href@noop {} {\bibfield  {journal} {\bibinfo
  {journal} {IEEE Transactions on Instrumentation and Measurement}\ }\textbf
  {\bibinfo {volume} {62}},\ \bibinfo {pages} {1556} (\bibinfo {year}
  {2013})}\BibitemShut {NoStop}%
\bibitem [{\citenamefont {Yamaguchi}\ \emph {et~al.}(2012)\citenamefont
  {Yamaguchi}, \citenamefont {Shiga}, \citenamefont {Nagano}, \citenamefont
  {Li}, \citenamefont {Ishijima}, \citenamefont {Hachisu}, \citenamefont
  {Kumagai},\ and\ \citenamefont {Ido}}]{Yamaguchi2012}%
  \BibitemOpen
  \bibfield  {author} {\bibinfo {author} {\bibfnamefont {A.}~\bibnamefont
  {Yamaguchi}}, \bibinfo {author} {\bibfnamefont {N.}~\bibnamefont {Shiga}},
  \bibinfo {author} {\bibfnamefont {S.}~\bibnamefont {Nagano}}, \bibinfo
  {author} {\bibfnamefont {Y.}~\bibnamefont {Li}}, \bibinfo {author}
  {\bibfnamefont {H.}~\bibnamefont {Ishijima}}, \bibinfo {author}
  {\bibfnamefont {H.}~\bibnamefont {Hachisu}}, \bibinfo {author} {\bibfnamefont
  {M.}~\bibnamefont {Kumagai}}, \ and\ \bibinfo {author} {\bibfnamefont
  {T.}~\bibnamefont {Ido}},\ }\href@noop {} {\bibfield  {journal} {\bibinfo
  {journal} {Applied Physics Express}\ }\textbf {\bibinfo {volume} {5}},\
  \bibinfo {pages} {022701} (\bibinfo {year} {2012})}\BibitemShut {NoStop}%
\bibitem [{\citenamefont {Nicolodi}\ \emph {et~al.}(2014)\citenamefont
  {Nicolodi}, \citenamefont {Argence}, \citenamefont {Zhang}, \citenamefont
  {Le~Targat}, \citenamefont {Santarelli},\ and\ \citenamefont
  {Le~Coq}}]{nicolodi2014spectral}%
  \BibitemOpen
  \bibfield  {author} {\bibinfo {author} {\bibfnamefont {D.}~\bibnamefont
  {Nicolodi}}, \bibinfo {author} {\bibfnamefont {B.}~\bibnamefont {Argence}},
  \bibinfo {author} {\bibfnamefont {W.}~\bibnamefont {Zhang}}, \bibinfo
  {author} {\bibfnamefont {R.}~\bibnamefont {Le~Targat}}, \bibinfo {author}
  {\bibfnamefont {G.}~\bibnamefont {Santarelli}}, \ and\ \bibinfo {author}
  {\bibfnamefont {Y.}~\bibnamefont {Le~Coq}},\ }\href@noop {} {\bibfield
  {journal} {\bibinfo  {journal} {Nature Photonics}\ }\textbf {\bibinfo
  {volume} {8}},\ \bibinfo {pages} {219} (\bibinfo {year} {2014})}\BibitemShut
  {NoStop}%
\bibitem [{\citenamefont {Scharnhorst}\ \emph {et~al.}(2015)\citenamefont
  {Scharnhorst}, \citenamefont {W{\"u}bbena}, \citenamefont {Hannig},
  \citenamefont {Jakobsen}, \citenamefont {Kramer}, \citenamefont {Leroux},\
  and\ \citenamefont {Schmidt}}]{Scharnhorst2015}%
  \BibitemOpen
  \bibfield  {author} {\bibinfo {author} {\bibfnamefont {N.}~\bibnamefont
  {Scharnhorst}}, \bibinfo {author} {\bibfnamefont {J.~B.}\ \bibnamefont
  {W{\"u}bbena}}, \bibinfo {author} {\bibfnamefont {S.}~\bibnamefont {Hannig}},
  \bibinfo {author} {\bibfnamefont {K.}~\bibnamefont {Jakobsen}}, \bibinfo
  {author} {\bibfnamefont {J.}~\bibnamefont {Kramer}}, \bibinfo {author}
  {\bibfnamefont {I.~D.}\ \bibnamefont {Leroux}}, \ and\ \bibinfo {author}
  {\bibfnamefont {P.~O.}\ \bibnamefont {Schmidt}},\ }\href@noop {} {\bibfield
  {journal} {\bibinfo  {journal} {Optics express}\ }\textbf {\bibinfo {volume}
  {23}},\ \bibinfo {pages} {19771} (\bibinfo {year} {2015})}\BibitemShut
  {NoStop}%
\bibitem [{\citenamefont {Su}\ \emph {et~al.}(2015)\citenamefont {Su},
  \citenamefont {Yan-Yi}, \citenamefont {Hai-Qin}, \citenamefont {Yuan},
  \citenamefont {Zhi-Yi},\ and\ \citenamefont {Long-Sheng}}]{Fang2015}%
  \BibitemOpen
  \bibfield  {author} {\bibinfo {author} {\bibfnamefont {F.}~\bibnamefont
  {Su}}, \bibinfo {author} {\bibfnamefont {J.}~\bibnamefont {Yan-Yi}}, \bibinfo
  {author} {\bibfnamefont {C.}~\bibnamefont {Hai-Qin}}, \bibinfo {author}
  {\bibfnamefont {Y.}~\bibnamefont {Yuan}}, \bibinfo {author} {\bibfnamefont
  {B.}~\bibnamefont {Zhi-Yi}}, \ and\ \bibinfo {author} {\bibfnamefont
  {M.}~\bibnamefont {Long-Sheng}},\ }\href@noop {} {\bibfield  {journal}
  {\bibinfo  {journal} {Chinese Physics B}\ }\textbf {\bibinfo {volume} {24}},\
  \bibinfo {pages} {074202} (\bibinfo {year} {2015})}\BibitemShut {NoStop}%
\bibitem [{\citenamefont {Akamatsu}\ \emph {et~al.}(2013)\citenamefont
  {Akamatsu}, \citenamefont {Inaba}, \citenamefont {Hosaka}, \citenamefont
  {Yasuda}, \citenamefont {Onae}, \citenamefont {Suzuyama}, \citenamefont
  {Amemiya},\ and\ \citenamefont {Hong}}]{akamatsu2013spectroscopy}%
  \BibitemOpen
  \bibfield  {author} {\bibinfo {author} {\bibfnamefont {D.}~\bibnamefont
  {Akamatsu}}, \bibinfo {author} {\bibfnamefont {H.}~\bibnamefont {Inaba}},
  \bibinfo {author} {\bibfnamefont {K.}~\bibnamefont {Hosaka}}, \bibinfo
  {author} {\bibfnamefont {M.}~\bibnamefont {Yasuda}}, \bibinfo {author}
  {\bibfnamefont {A.}~\bibnamefont {Onae}}, \bibinfo {author} {\bibfnamefont
  {T.}~\bibnamefont {Suzuyama}}, \bibinfo {author} {\bibfnamefont
  {M.}~\bibnamefont {Amemiya}}, \ and\ \bibinfo {author} {\bibfnamefont
  {F.-L.}\ \bibnamefont {Hong}},\ }\href@noop {} {\bibfield  {journal}
  {\bibinfo  {journal} {Applied Physics Express}\ }\textbf {\bibinfo {volume}
  {7}},\ \bibinfo {pages} {012401} (\bibinfo {year} {2013})}\BibitemShut
  {NoStop}%
\bibitem [{\citenamefont {Inaba}\ \emph {et~al.}(2013)\citenamefont {Inaba},
  \citenamefont {Hosaka}, \citenamefont {Yasuda}, \citenamefont {Nakajima},
  \citenamefont {Iwakuni}, \citenamefont {Akamatsu}, \citenamefont {Okubo},
  \citenamefont {Kohno}, \citenamefont {Onae},\ and\ \citenamefont
  {Hong}}]{Inaba2013}%
  \BibitemOpen
  \bibfield  {author} {\bibinfo {author} {\bibfnamefont {H.}~\bibnamefont
  {Inaba}}, \bibinfo {author} {\bibfnamefont {K.}~\bibnamefont {Hosaka}},
  \bibinfo {author} {\bibfnamefont {M.}~\bibnamefont {Yasuda}}, \bibinfo
  {author} {\bibfnamefont {Y.}~\bibnamefont {Nakajima}}, \bibinfo {author}
  {\bibfnamefont {K.}~\bibnamefont {Iwakuni}}, \bibinfo {author} {\bibfnamefont
  {D.}~\bibnamefont {Akamatsu}}, \bibinfo {author} {\bibfnamefont
  {S.}~\bibnamefont {Okubo}}, \bibinfo {author} {\bibfnamefont
  {T.}~\bibnamefont {Kohno}}, \bibinfo {author} {\bibfnamefont
  {A.}~\bibnamefont {Onae}}, \ and\ \bibinfo {author} {\bibfnamefont {F.-L.}\
  \bibnamefont {Hong}},\ }\href@noop {} {\bibfield  {journal} {\bibinfo
  {journal} {Optics express}\ }\textbf {\bibinfo {volume} {21}},\ \bibinfo
  {pages} {7891} (\bibinfo {year} {2013})}\BibitemShut {NoStop}%
\bibitem [{\citenamefont {Akamatsu}\ \emph {et~al.}(2012)\citenamefont
  {Akamatsu}, \citenamefont {Nakajima}, \citenamefont {Inaba}, \citenamefont
  {Hosaka}, \citenamefont {Yasuda}, \citenamefont {Onae},\ and\ \citenamefont
  {Hong}}]{Akamatsu:12}%
  \BibitemOpen
  \bibfield  {author} {\bibinfo {author} {\bibfnamefont {D.}~\bibnamefont
  {Akamatsu}}, \bibinfo {author} {\bibfnamefont {Y.}~\bibnamefont {Nakajima}},
  \bibinfo {author} {\bibfnamefont {H.}~\bibnamefont {Inaba}}, \bibinfo
  {author} {\bibfnamefont {K.}~\bibnamefont {Hosaka}}, \bibinfo {author}
  {\bibfnamefont {M.}~\bibnamefont {Yasuda}}, \bibinfo {author} {\bibfnamefont
  {A.}~\bibnamefont {Onae}}, \ and\ \bibinfo {author} {\bibfnamefont {F.-L.}\
  \bibnamefont {Hong}},\ }\href@noop {} {\bibfield  {journal} {\bibinfo
  {journal} {Optics express}\ }\textbf {\bibinfo {volume} {20}},\ \bibinfo
  {pages} {16010} (\bibinfo {year} {2012})}\BibitemShut {NoStop}%
\bibitem [{\citenamefont {Yamanaka}\ \emph {et~al.}(2015)\citenamefont
  {Yamanaka}, \citenamefont {Ohmae}, \citenamefont {Ushijima}, \citenamefont
  {Takamoto},\ and\ \citenamefont {Katori}}]{yamanaka2015frequency}%
  \BibitemOpen
  \bibfield  {author} {\bibinfo {author} {\bibfnamefont {K.}~\bibnamefont
  {Yamanaka}}, \bibinfo {author} {\bibfnamefont {N.}~\bibnamefont {Ohmae}},
  \bibinfo {author} {\bibfnamefont {I.}~\bibnamefont {Ushijima}}, \bibinfo
  {author} {\bibfnamefont {M.}~\bibnamefont {Takamoto}}, \ and\ \bibinfo
  {author} {\bibfnamefont {H.}~\bibnamefont {Katori}},\ }\href@noop {}
  {\bibfield  {journal} {\bibinfo  {journal} {Physical review letters}\
  }\textbf {\bibinfo {volume} {114}},\ \bibinfo {pages} {230801} (\bibinfo
  {year} {2015})}\BibitemShut {NoStop}%
\bibitem [{\citenamefont {Nemitz}\ \emph {et~al.}(2016)\citenamefont {Nemitz},
  \citenamefont {Ohkubo}, \citenamefont {Takamoto}, \citenamefont {Ushijima},
  \citenamefont {Das}, \citenamefont {Ohmae},\ and\ \citenamefont
  {Katori}}]{nemitz2016frequency}%
  \BibitemOpen
  \bibfield  {author} {\bibinfo {author} {\bibfnamefont {N.}~\bibnamefont
  {Nemitz}}, \bibinfo {author} {\bibfnamefont {T.}~\bibnamefont {Ohkubo}},
  \bibinfo {author} {\bibfnamefont {M.}~\bibnamefont {Takamoto}}, \bibinfo
  {author} {\bibfnamefont {I.}~\bibnamefont {Ushijima}}, \bibinfo {author}
  {\bibfnamefont {M.}~\bibnamefont {Das}}, \bibinfo {author} {\bibfnamefont
  {N.}~\bibnamefont {Ohmae}}, \ and\ \bibinfo {author} {\bibfnamefont
  {H.}~\bibnamefont {Katori}},\ }\href@noop {} {\bibfield  {journal} {\bibinfo
  {journal} {Nature Photonics}\ }\textbf {\bibinfo {volume} {10}},\ \bibinfo
  {pages} {258} (\bibinfo {year} {2016})}\BibitemShut {NoStop}%
\bibitem [{\citenamefont {Yamaguchi}\ \emph {et~al.}(2011)\citenamefont
  {Yamaguchi}, \citenamefont {Fujieda}, \citenamefont {Kumagai}, \citenamefont
  {Hachisu}, \citenamefont {Nagano}, \citenamefont {Li}, \citenamefont {Ido},
  \citenamefont {Takano}, \citenamefont {Takamoto},\ and\ \citenamefont
  {Katori}}]{yamaguchi2011direct}%
  \BibitemOpen
  \bibfield  {author} {\bibinfo {author} {\bibfnamefont {A.}~\bibnamefont
  {Yamaguchi}}, \bibinfo {author} {\bibfnamefont {M.}~\bibnamefont {Fujieda}},
  \bibinfo {author} {\bibfnamefont {M.}~\bibnamefont {Kumagai}}, \bibinfo
  {author} {\bibfnamefont {H.}~\bibnamefont {Hachisu}}, \bibinfo {author}
  {\bibfnamefont {S.}~\bibnamefont {Nagano}}, \bibinfo {author} {\bibfnamefont
  {Y.}~\bibnamefont {Li}}, \bibinfo {author} {\bibfnamefont {T.}~\bibnamefont
  {Ido}}, \bibinfo {author} {\bibfnamefont {T.}~\bibnamefont {Takano}},
  \bibinfo {author} {\bibfnamefont {M.}~\bibnamefont {Takamoto}}, \ and\
  \bibinfo {author} {\bibfnamefont {H.}~\bibnamefont {Katori}},\ }\href@noop {}
  {\bibfield  {journal} {\bibinfo  {journal} {Applied physics express}\
  }\textbf {\bibinfo {volume} {4}},\ \bibinfo {pages} {082203} (\bibinfo {year}
  {2011})}\BibitemShut {NoStop}%
\bibitem [{\citenamefont {Ye}\ and\ \citenamefont {Cundiff}(2005)}]{Ye2005}%
  \BibitemOpen
  \bibfield  {author} {\bibinfo {author} {\bibfnamefont {J.}~\bibnamefont
  {Ye}}\ and\ \bibinfo {author} {\bibfnamefont {S.~T.}\ \bibnamefont
  {Cundiff}},\ }\href@noop {} {\emph {\bibinfo {title} {Femtosecond optical
  frequency comb: principle, operation and applications}}}\ (\bibinfo
  {publisher} {Springer Science \& Business Media},\ \bibinfo {year}
  {2005})\BibitemShut {NoStop}%
\bibitem [{\citenamefont {Newbury}\ and\ \citenamefont
  {Swann}(2007)}]{newbury2007low}%
  \BibitemOpen
  \bibfield  {author} {\bibinfo {author} {\bibfnamefont {N.~R.}\ \bibnamefont
  {Newbury}}\ and\ \bibinfo {author} {\bibfnamefont {W.~C.}\ \bibnamefont
  {Swann}},\ }\href@noop {} {\bibfield  {journal} {\bibinfo  {journal} {JOSA
  B}\ }\textbf {\bibinfo {volume} {24}},\ \bibinfo {pages} {1756} (\bibinfo
  {year} {2007})}\BibitemShut {NoStop}%
\bibitem [{\citenamefont {Stenger}\ \emph {et~al.}(2002)\citenamefont
  {Stenger}, \citenamefont {Schnatz}, \citenamefont {Tamm},\ and\ \citenamefont
  {Telle}}]{stenger2002ultraprecise}%
  \BibitemOpen
  \bibfield  {author} {\bibinfo {author} {\bibfnamefont {J.}~\bibnamefont
  {Stenger}}, \bibinfo {author} {\bibfnamefont {H.}~\bibnamefont {Schnatz}},
  \bibinfo {author} {\bibfnamefont {C.}~\bibnamefont {Tamm}}, \ and\ \bibinfo
  {author} {\bibfnamefont {H.~R.}\ \bibnamefont {Telle}},\ }\href@noop {}
  {\bibfield  {journal} {\bibinfo  {journal} {Physical review letters}\
  }\textbf {\bibinfo {volume} {88}},\ \bibinfo {pages} {073601} (\bibinfo
  {year} {2002})}\BibitemShut {NoStop}%
\bibitem [{\citenamefont {Grosche}\ \emph {et~al.}(2008)\citenamefont
  {Grosche}, \citenamefont {Lipphardt},\ and\ \citenamefont
  {Schnatz}}]{Grosche2008}%
  \BibitemOpen
  \bibfield  {author} {\bibinfo {author} {\bibfnamefont {G.}~\bibnamefont
  {Grosche}}, \bibinfo {author} {\bibfnamefont {B.}~\bibnamefont {Lipphardt}},
  \ and\ \bibinfo {author} {\bibfnamefont {H.}~\bibnamefont {Schnatz}},\ }\href
  {\doibase 10.1140/epjd/e2008-00065-7} {\bibfield  {journal} {\bibinfo
  {journal} {The European Physical Journal D}\ }\textbf {\bibinfo {volume}
  {48}},\ \bibinfo {pages} {27} (\bibinfo {year} {2008})}\BibitemShut {NoStop}%
\bibitem [{\citenamefont {Fang}\ \emph {et~al.}(2013)\citenamefont {Fang},
  \citenamefont {Chen}, \citenamefont {Wang}, \citenamefont {Jiang},
  \citenamefont {Bi},\ and\ \citenamefont {Ma}}]{Fang2013Optical}%
  \BibitemOpen
  \bibfield  {author} {\bibinfo {author} {\bibfnamefont {S.}~\bibnamefont
  {Fang}}, \bibinfo {author} {\bibfnamefont {H.}~\bibnamefont {Chen}}, \bibinfo
  {author} {\bibfnamefont {T.}~\bibnamefont {Wang}}, \bibinfo {author}
  {\bibfnamefont {Y.}~\bibnamefont {Jiang}}, \bibinfo {author} {\bibfnamefont
  {Z.}~\bibnamefont {Bi}}, \ and\ \bibinfo {author} {\bibfnamefont
  {L.}~\bibnamefont {Ma}},\ }\href {\doibase 10.1063/1.4809736} {\bibfield
  {journal} {\bibinfo  {journal} {Applied Physics Letters}\ }\textbf {\bibinfo
  {volume} {102}},\ \bibinfo {pages} {231118} (\bibinfo {year} {2013})},\
  \Eprint {http://arxiv.org/abs/https://doi.org/10.1063/1.4809736}
  {https://doi.org/10.1063/1.4809736} \BibitemShut {NoStop}%
\bibitem [{\citenamefont {Nakajima}\ \emph {et~al.}(2010)\citenamefont
  {Nakajima}, \citenamefont {Inaba}, \citenamefont {Hosaka}, \citenamefont
  {Minoshima}, \citenamefont {Onae}, \citenamefont {Yasuda}, \citenamefont
  {Kohno}, \citenamefont {Kawato}, \citenamefont {Kobayashi}, \citenamefont
  {Katsuyama} \emph {et~al.}}]{nakajima2010multi}%
  \BibitemOpen
  \bibfield  {author} {\bibinfo {author} {\bibfnamefont {Y.}~\bibnamefont
  {Nakajima}}, \bibinfo {author} {\bibfnamefont {H.}~\bibnamefont {Inaba}},
  \bibinfo {author} {\bibfnamefont {K.}~\bibnamefont {Hosaka}}, \bibinfo
  {author} {\bibfnamefont {K.}~\bibnamefont {Minoshima}}, \bibinfo {author}
  {\bibfnamefont {A.}~\bibnamefont {Onae}}, \bibinfo {author} {\bibfnamefont
  {M.}~\bibnamefont {Yasuda}}, \bibinfo {author} {\bibfnamefont
  {T.}~\bibnamefont {Kohno}}, \bibinfo {author} {\bibfnamefont
  {S.}~\bibnamefont {Kawato}}, \bibinfo {author} {\bibfnamefont
  {T.}~\bibnamefont {Kobayashi}}, \bibinfo {author} {\bibfnamefont
  {T.}~\bibnamefont {Katsuyama}},  \emph {et~al.},\ }\href@noop {} {\bibfield
  {journal} {\bibinfo  {journal} {Optics Express}\ }\textbf {\bibinfo {volume}
  {18}},\ \bibinfo {pages} {1667} (\bibinfo {year} {2010})}\BibitemShut
  {NoStop}%
\bibitem [{\citenamefont {Ohmae}\ \emph {et~al.}(2017)\citenamefont {Ohmae},
  \citenamefont {Kuse}, \citenamefont {Fermann},\ and\ \citenamefont
  {Katori}}]{ohmae2017all}%
  \BibitemOpen
  \bibfield  {author} {\bibinfo {author} {\bibfnamefont {N.}~\bibnamefont
  {Ohmae}}, \bibinfo {author} {\bibfnamefont {N.}~\bibnamefont {Kuse}},
  \bibinfo {author} {\bibfnamefont {M.~E.}\ \bibnamefont {Fermann}}, \ and\
  \bibinfo {author} {\bibfnamefont {H.}~\bibnamefont {Katori}},\ }\href@noop {}
  {\bibfield  {journal} {\bibinfo  {journal} {Applied Physics Express}\
  }\textbf {\bibinfo {volume} {10}},\ \bibinfo {pages} {062503} (\bibinfo
  {year} {2017})}\BibitemShut {NoStop}%
\bibitem [{\citenamefont {Angrisani}\ \emph {et~al.}(2001)\citenamefont
  {Angrisani}, \citenamefont {D'Apuzzo},\ and\ \citenamefont
  {D'Arco}}]{angrisani2001digital}%
  \BibitemOpen
  \bibfield  {author} {\bibinfo {author} {\bibfnamefont {L.}~\bibnamefont
  {Angrisani}}, \bibinfo {author} {\bibfnamefont {M.}~\bibnamefont {D'Apuzzo}},
  \ and\ \bibinfo {author} {\bibfnamefont {M.}~\bibnamefont {D'Arco}},\
  }\href@noop {} {\bibfield  {journal} {\bibinfo  {journal} {IEEE Transactions
  on instrumentation and measurement}\ }\textbf {\bibinfo {volume} {50}},\
  \bibinfo {pages} {930} (\bibinfo {year} {2001})}\BibitemShut {NoStop}%
\bibitem [{\citenamefont {Akerman}\ \emph {et~al.}(2012)\citenamefont
  {Akerman}, \citenamefont {Glickman}, \citenamefont {Kotler}, \citenamefont
  {Keselman},\ and\ \citenamefont {Ozeri}}]{Akerman2012}%
  \BibitemOpen
  \bibfield  {author} {\bibinfo {author} {\bibfnamefont {N.}~\bibnamefont
  {Akerman}}, \bibinfo {author} {\bibfnamefont {Y.}~\bibnamefont {Glickman}},
  \bibinfo {author} {\bibfnamefont {S.}~\bibnamefont {Kotler}}, \bibinfo
  {author} {\bibfnamefont {A.}~\bibnamefont {Keselman}}, \ and\ \bibinfo
  {author} {\bibfnamefont {R.}~\bibnamefont {Ozeri}},\ }\href {\doibase
  10.1007/s00340-011-4807-6} {\bibfield  {journal} {\bibinfo  {journal}
  {Applied Physics B}\ }\textbf {\bibinfo {volume} {107}},\ \bibinfo {pages}
  {1167} (\bibinfo {year} {2012})}\BibitemShut {NoStop}%
\bibitem [{\citenamefont {Akerman}\ \emph {et~al.}(2015)\citenamefont
  {Akerman}, \citenamefont {Navon}, \citenamefont {Kotler}, \citenamefont
  {Glickman},\ and\ \citenamefont {Ozeri}}]{Akerman2015}%
  \BibitemOpen
  \bibfield  {author} {\bibinfo {author} {\bibfnamefont {N.}~\bibnamefont
  {Akerman}}, \bibinfo {author} {\bibfnamefont {N.}~\bibnamefont {Navon}},
  \bibinfo {author} {\bibfnamefont {S.}~\bibnamefont {Kotler}}, \bibinfo
  {author} {\bibfnamefont {Y.}~\bibnamefont {Glickman}}, \ and\ \bibinfo
  {author} {\bibfnamefont {R.}~\bibnamefont {Ozeri}},\ }\href@noop {}
  {\bibfield  {journal} {\bibinfo  {journal} {New Journal of Physics}\ }\textbf
  {\bibinfo {volume} {17}},\ \bibinfo {pages} {113060} (\bibinfo {year}
  {2015})}\BibitemShut {NoStop}%
\bibitem [{\citenamefont {Agarwal}(1978)}]{agarwal1978quantum}%
  \BibitemOpen
  \bibfield  {author} {\bibinfo {author} {\bibfnamefont {G.}~\bibnamefont
  {Agarwal}},\ }\href@noop {} {\bibfield  {journal} {\bibinfo  {journal}
  {Physical Review A}\ }\textbf {\bibinfo {volume} {18}},\ \bibinfo {pages}
  {1490} (\bibinfo {year} {1978})}\BibitemShut {NoStop}%
\end{thebibliography}%

\end{document}